\documentclass[]{aastex631}
\usepackage{amsmath}
\usepackage{pifont}

\newcommand{\eb}{\begin{equation}}
\newcommand{\ee}{\end{equation}}

\definecolor{rkka}{RGB}{180,12,15}
\definecolor{nsgreen}{rgb}{0.1,0.5,0.1}

\shorttitle{Distributions of wide stellar binary orbits}
\shortauthors{Makarov}

\begin{document}

\title{Distributions of wide binary stars in theory and in Gaia data:\\
II. Reconstruction of sample probability density of true orbit sizes}

\correspondingauthor{Valeri V. Makarov}
\email{valeri.makarov@gmail.com}

\author[0000-0003-2336-7887]{Valeri V. Makarov}
\affiliation{U.S. Naval Observatory, 3450 Massachusetts Ave NW, Washington, DC 20392-5420, USA}

\begin{abstract}
Wide binary stars are important for testing alternative models of gravitation in the weak-field regime and understanding the statistical outcomes of dynamical interactions in the general Galactic field. The Gaia mission's collection of weakly bound pairs of stars offers a unique opportunity to estimate the rate of survivors at separations above 7 KAU, where non-Newtonian components of gravitation may become important. The available Gaia-based catalogs of resolved binaries provide the projected angular separation between the components, while the physical semimajor axis is the parameter of interest. The problem of reconstructing the distribution of orbit sizes is complex and ill-posed, because the observed apparent separations are defined by a number of underlying physical parameters including semimajor axis, eccentricity, orbit orientation, and orbital phase. Methods of inverse reconstruction of the marginal distribution of semimajor axes from the available data are proposed and implemented with two different strategies, namely, a direct Monte Carlo mapping and inverse filtering with impulse updates without regularization. We find rather similar results from the two methods for the outlying tail of the distribution, suggesting that the rate of orbit sizes is a shallow declining function on a logarithmic scale. A finite rate of extremely wide binaries is implied.

 \end{abstract}

\section{Introduction} \label{int.sec}
The observed statistical distributions of directly measurable parameters of wide, resolved stellar binaries and hierarchical multiples, such as projected separation, position angle, and relative velocity,  are complex transformations of the underlying population distributions of basic physical parameters, including mass, semimajor axis, eccentricity, and the Euler angles of geometric orientation in a fixed inertial reference frame. The basic distributions of orbit size and eccentricity, and possible relations between them, are of special interest for cosmogony of such systems, finding evidence of long-term intrinsic evolution including the effects of tidal acceleration in the Galactic potential, interaction with the ambient stellar field, and, potentially, testing alternative concepts of gravitation in the weak-field regime. 

The binding energy of wide binary stars is low, which makes them vulnerable to external perturbations. They should be inherently less stable than tight (also called hard) binaries, which require much larger inputs of external kinetic energy to become disassociated. The most probable outcome of an encounter of a wide, weakly bound binary with a chance perturber at a sufficiently small impact parameter is the disruption of the initial system \citep{1975AJ.....80..809H, 1975MNRAS.173..729H, 1987ApJ...312..367W}. Given that many of field stars were formed in denser open cluster environments, where such interactions are significantly more frequent, and the process of cluster dispersal involves violent dynamical encounters with massive and hard binaries in the core \citep{2001MNRAS.321..199P}, the presence of soft binaries in the old disk population is a theoretical challenge.

It is traditional, as well as practically convenient, to consider in this analysis the probability density of $\lambda=\log(s)$, where 
\eb 
s=\rho/\varpi,
\label{rho.eq}
\ee 
$\rho$ being the angular separation between the components in mas, $\varpi$ the measured (or assumed) parallax in mas, and $s$ is  the observed separation expressed in AU.
The convention of expressing the probability density in decibels instead of the actual value follows the pioneering works by \"Opik and his early hypothesis that the tail of this distribution follows PDF$(\lambda)=\;$const \citep{1924PTarO..25f...1O}. This model, corresponding to a slowly declining rate of binaries proportional to $1/s$, was empirically derived using very limited catalogs of nearby systems with numerous selection biases and technical features. Almost a century later, additional studies in this direction suffered from similar problems of small-number statistics and limited range of  separation \citep[e.g.,][]{2002A&A...382...92S}, which generally confirmed this proposal. \citet{2007AJ....133..889L}, using a sample of just 521 Hipparcos pairs with common proper motions in the northern hemisphere, deduced that \"Opik's law is confirmed for separations up to 4 KAU, but the decline in the rate of binaries becomes faster with PDF$[s]\propto s^{-(1.6\pm 0.1)}$ for wider separations. An even smaller Hipparcos-based sample of extremely wide pairs was investigated by \citet{2008ApJ...687..566M}, revealing the presence of separations up to 1 pc and an apparent lack of Galactic halo systems. \citet{2004ApJ...601..289C} suggested that the disk and halo pairs follow similar power-law probability functions, which can be an argument in favor of their common formation mechanism. 

The actual parameter of interest is
$a$, the semimajor axis (also loosely called true orbital separation) in AU, not the observed projected separation. In some papers, these two parameters are tacitly equated \citep[e.g.,][]{2012AASP....2...23C}, which is not accurate. Indeed, the mapping of the true semimajor axis $a$ onto the instantaneous orbital distance $r$ and further to the observed projected separation $s$ is complex and nonlinear, involving separate draws from the sample distributions of several orbital parameters, including the eccentricity. \citet{2007A&A...474...77K} took a deeper approach by forward modeling of $s$ for binaries in the ScoCen OB2 young association, assuming a thermal marginal distribution of eccentricity (without any supporting evidence for this) and a number of other propositions. They used the Kolmogorov-Smirnov test and the goodness-of-fit metric ($p$-value) and arrived at a rough estimate that PDF$(a)\propto a^{-\gamma}$ with estimated values of $\gamma$ between $-1.2$ and $-0.9$. This result appears to be very roughly consistent with \"Opik's model ($\gamma=-1$) extended to the space of true semimajor axes.

This study extends the analysis of the sample distribution of $a$ to an all-sky sample of Gaia-detected resolved binaries totaling $\sim 10^5$ pairs after careful filtering and vetting. We will connect the observable separation $s$ and the true orbital axis $a$ in a statistical simulation by introducing an intermediate relative projection factor, which is only slightly sensitive to the exact shape of the eccentricity distribution. The questions we want to address are:
\begin{enumerate}
    \item How far does the general field population of soft binaries stretch in terms of the orbit sizes?
    \item Can we find a more accurate distribution law of orbit sizes for the widest systems beyond the previously used assumption of a uniform power-law?
\end{enumerate}

The orbital parameters involved in the Monte Carlo mapping of $a$ to $s$ are mostly well known from the assumed isotropy of orbit orientation in space and orbit phase in time (mean anomaly). The only exception is the orbital eccentricity. The intrinsic distribution of eccentricity has been a subject of intensive investigation for many decades \citep[e.g.,][]{2010ApJS..190....1R, 2017ApJS..230...15M, 2012AJ....144...54G, 2013ApJ...779...30G, 2020MNRAS.496..987T}. The sample distribution may be the nontrivial result of the dominant formation scenarios (such as core fragmentation with migration and friction-assisted capture) and the subsequent dynamical interaction with other stars and binary systems. Since most stars are formed in open clusters and star-forming associations, the dynamical environment and binary statistics are of special significance for these aggregates of stars. The disruption/capture events may combine in sequential chain scenarios, reshaping the primordial distribution of eccentricity \citep{2012ApJ...750...83P}. The expected net result for surviving old systems in the field  is the so-called thermal (${\rm PDF}(e)=2\,e$) or superthermal (pileup toward $e=1$) distributions of eccentricity. In this study, a single power-law distribution function is assumed based on the pervious analysis of the same data set \citep{2025AJ....169..113M}. This choice has a limited effect on the conclusions because of the low sensitivity of the projection coefficients to the exact shape of PDF$(e)$.

\section{The Gaia-based data sample}
\label{gaia.sec}
The primary data sample used in this study is identical to the collection of resolved binary systems in \citep{2025AJ....169..113M}. It is derived from the large, all-sky catalog of candidate binary systems (1.3 million pairs) by \citet{2021MNRAS.506.2269E}. That catalog is based on the data from the third Gaia mission release \citep{2016A&A...595A...1G, 2023A&A...674A...1G}. Stringent additional filters have been applied to remove the possible contaminants due to chance alignments of field stars and members of stellar clusters and associations. One of the filters utilizes the cataloged chance alignment probability parameter ${\cal R}$, where an aggressive cut is made at ${\cal R}<0.01$ (the authors recommend a threshold value of 0.1). Numerous distant pairs with the mean parallax below 4 mas were removed because their astrometric determinations are less reliable due to unresolved binarity \citep{2021A&A...649A...5F}. All pairs with angular separations below $2\arcsec$ were also discarded, because the quality of data may be impaired by the limited angular resolution of Gaia. The cut does not affect the objectives of this study, which is focused on wider binaries, because, as is shown in this paper, clipping a small fraction of cataloged pairs with $\rho<2\arcsec$ in combination with the filter $\varpi>4$ mas does not significantly change the reconstructed $a$-distribution at $a>7$ KAU. An additional filter on the Gaia-specific {\tt ruwe} parameter, which quantifies the degree of compliance of each source with the 5- or 6-parameter astrometric model, is meant to remove remaining wide pairs that may represent hierarchical multiple systems, although those are less dangerous in the context of this paper. The applied filter is {\tt ruwe}$<1.3$ for both components.

The number of binaries is drastically reduced with these filters from an initial 1.3 million to 103,169. For comparison, the earlier analysis of the apparent separations for three roughly segregated kinematic components of the Galaxy by \citet{Tian_2020} used an almost 8 times larger sample. The smallest apparent separation $s$ in our working sample is 25.04 AU, with only 1\% of the sample shorter than 193.8 AU. The main reason for the lack of tight binaries in the source sample is the limited angular resolution of Gaia astrometry \citep{2021A&A...649A...5F}. The maximum projected separation is 161498 AU, with 1\% of the filtered sample above 24461 AU. This indicates the presence of pairs with $a>7$ KAU, since $a$ cannot be smaller than $s/2$.

\section{Assumptions and approximations} \label{ass.sec}
The observed histogram of projected separations between the components of Gaia binaries depends in a complex, nonlinear way on the sample distribution of orbital eccentricity. Here, we consistently use the specific power-law model for the phase-density  of binaries at a fixed energy $\cal E$ suggested by \citet{2025AJ....169..113M}.
The cumulative distribution function (CDF) and the probability density function (PDF) of eccentricity in that model are
\begin{eqnarray}
    {\rm CDF}(e)&=&1-(1-e^2)^{1-\alpha'/2}\nonumber \\
    {\rm PDF}(e)&=&(2-\alpha')e(1-e^2)^{-\alpha'/2}.
    \label{pl.eq}
\end{eqnarray}
The so-called thermal distribution of eccentricity, ${\rm PDF}(e)=2e$ is a particular case of this model with $\alpha'=0$. It has received special attention in the literature because of the early theoretical works, where a physical origin was proposed based on the assumed Boltzmann distribution law of orbital energy \citep{1919MNRAS..79..408J} or, more generally, on the assumption of a flat dependence of the 6D phase density of binary stars on all Delaunay elements except $L$ \citep{1937AZh....14..207A}. More recent analyses for smaller samples of resolved binaries \citep{2016MNRAS.456.2070T, 2022MNRAS.512.3383H} determined that the wider population of binary stars follows a ``superthermal" distribution law, with a concave, fast-rising density toward the marginal value $e=1$. Closer and moderately separated binaries 
are more consistent with peaked Rayleigh-type distributions \citep{2025ApJ...982L..34W}. A transition between these dynamically distinct classes takes place at approximately 1 KAU \citep{2017ApJS..230...15M}.

\begin{figure*}
    \includegraphics[width=0.33 \textwidth]{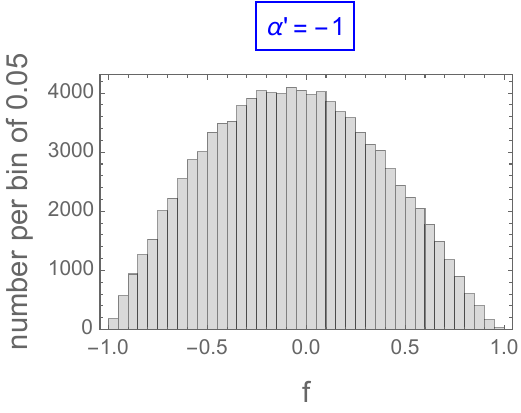}
    \includegraphics[width=0.33 \textwidth]{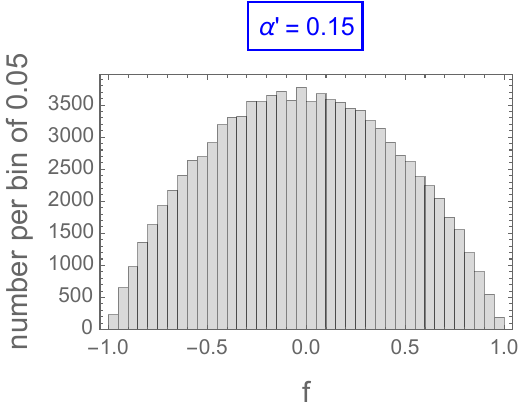}
    \includegraphics[width=0.33 \textwidth]{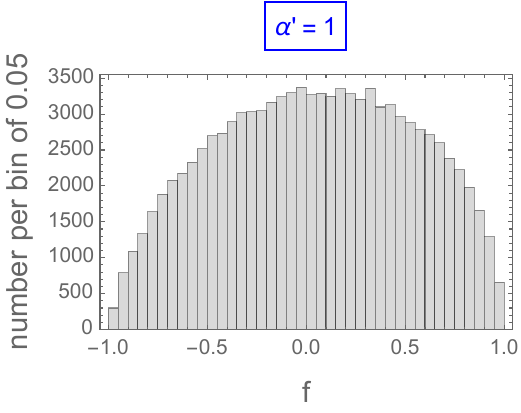}
    \caption{Monte Carlo-simulated histograms of $f$, which is the fractional projection factor given by Eq. \ref{f.eq}, for three cases of power-law eccentricity distributions, $\alpha'=-1$ (left),  $\alpha'=+0.15$ (center), and $\alpha'=+1$ (right).}
    \label{f.fig}
\end{figure*}

Let us define the fractional projection factor, which is convenient to use as an intermediate statistical parameter linking the observed separation in the sky plane $s$ and the true semimajor axis $a$ in 3D space:
\eb 
f\equiv \frac{s-a}{a}.
\label{f.eq}
\ee 
For a given pair of stars, this value depends on a number of orbital parameters, including eccentricity, instantaneous mean anomaly $M$ at the time of measurement, Euler angles of orientation in the observer's frame $\Omega$, $\omega$, and $i$. The PDF of $f$ is independent of $a$, as long as we can assume that the distribution of eccentricity is homomorphic for any fixed $a$. Indeed, the geometrical projection factors are scale-invariant, while the distributions of the Euler angles and $M$ are known a priori for an isotropic orientation of orbital momenta. Within these acceptable assumptions, we can now compute the distributions of $f$ for a specific power-law distribution of $e$ from Eq. \ref{pl.eq} using Monte Carlo simulations of the five orbital elements. The results with $10^5$ Monte Carlo trials for three selected values of $\alpha'$ are shown in Fig. \ref{f.fig}.

Even a brief look at the simulated distributions reveals a remarkably low sensitivity of $f$ to the shape of PDF$[e]$. Within the range of interest for $\alpha'$, all these functions are convex and fairly symmetric with modes in the vicinity of 0. Indeed, the computed mean $f$ values are $-0.058$, $-0.013$, and $+0.043$ for $\alpha'=-1$, $+0.15$, and $+1$, respectively. The corresponding median $f$ values are $-0.066$, $-0.017$, and $+0.050$. Thus, statistically speaking, the projection and eccentricity effects change the average separations for a fixed $a$ by only several percent, and the best-fitting, empirically estimated sample value $\alpha'=+0.15$ corresponds to a transformation that makes the projected separations slightly shorter on average than the true orbit sizes. This property explains why we find the empirical distribution of $s$ to be of relatively low importance for fitting the intrinsic distribution of eccentricity. Indeed, changing the parameter $\alpha'$ in a wide range is barely reflected in the shape of the simulated $s$-distribution\footnote{For superthermal distributions of eccentricity ($\alpha'>0$), the $f$-histograms lean toward positive values, which is intuitively clear. With orbital eccentricity piling up at 1, an increasing fraction of binaries are observed closer to their apoastrons, where the projected separation can be larger than $a$ by up to a factor of 2.}. On the other hand, this is an asset for the main objective of this study, because we can reconstruct the PDF$(a)$ for fixed $\alpha'=+0.15$ and claim that the result is robust with respect to moderate deviations from the accepted eccentricity model.

\section{Reconstruction of orbit separation by direct statistical mapping}
\label{di.sec}
We note that the numerically generated distributions of the projection factor $f$ in Eq. \ref{f.eq} allow us to statistically map $a\rightarrow s$ in Monte Carlo simulations by simply multiplying a fixed $a$ by randomly generated $1+f$. This mapping is numerically stable and free of singularities. The direct mapping $s\rightarrow a$ for each data point is also available by dividing a fixed (observed) $s$ by $1+f$. This possibility is realized via direct Monte Carlo simulations (called direct mapping, or DM) described in this Section.

\begin{figure*}
    \includegraphics[width=0.47 \textwidth]{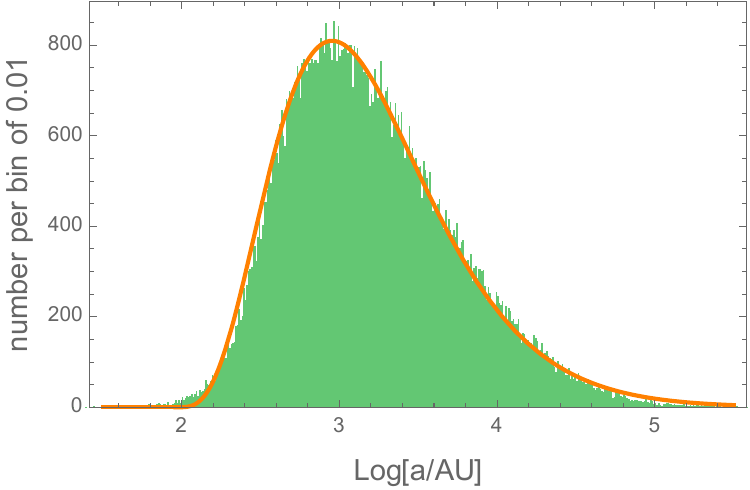}
    \caption{Distribution (histogram in green) of the logarithms of semimajor axes between the components of resolved binary systems in Gaia DR3 from the catalog by \citet{2021MNRAS.506.2269E} by direct Monte Carlo mapping. The orange line is the numerically fitted analytical distribution given by Eq. \ref{gamma1.eq}. }
    \label{di.fig}
\end{figure*}

For each value of $s$, a random value $f$ is generated from the previously computed distribution corresponding to an adopted power index $\alpha'$ (Fig. \ref{f.fig}). There is no need to represent the distribution of $f$ by a certain analytical form for this operation, because a sufficiently large simulated sample of $f$-values (I used $3\times 10^5$ trials) can be treated in a nonparametric way as an empirical distribution law\footnote{In Wolfram Mathematica, {\tt RandomVariate[EmpiricalDistribution[f],Length$@$s]} can be used.}. This is sufficient to generate a random realization of $a$ corresponding to the input value $s$. The numerical operation is fast, and a large number of trials can be generated if necessary. Fig. \ref{di.fig} shows the resulting MC-generated histogram for the ancillary parameter
\eb 
l=\log_{10}(a), \label{l.eq}
\ee 
where $a$ is expressed in AU. We find this substitute to be practically convenient, because the asymmetric and sharp-crested distribution of $s$ (see, e.g., Fig. 2 in \citet{2025AJ....169..113M}) corresponds to a much better behaved PDF in $l$ and is more amenable to accurate analytical representation. Indeed, a numerically fitted distribution law
\eb 
{\rm PDF}(l)=\Gamma[4.99,0.25](l-1.96),
\label{gamma1.eq}
\ee 
where $\Gamma$ stands for the Gamma distribution with shape 4.99 and scale 0.25 shifted by an offset of 1.96, faithfully represents the shape of the generated histogram shown in Fig. \ref{di.fig}. The largest deviations are seen for the closer subpopulation of Gaia binaries with $a<1$ KAU, which is inconsequential for the present study. It should be noted that the empirical fit pertains to the given filtered sample and does not extrapolate to the general population of field binaries. We also note that the fit overestimates the rate of binaries at $l>4.4$.

\begin{figure*}
    \includegraphics[width=0.47 \textwidth]{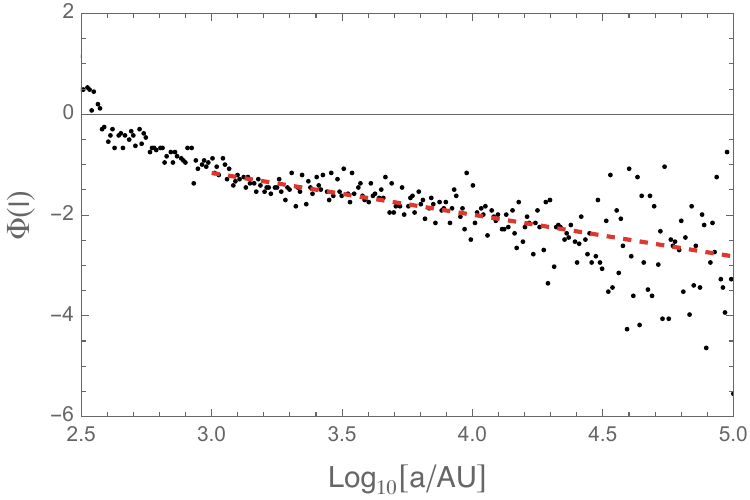}
    \includegraphics[width=0.47 \textwidth]{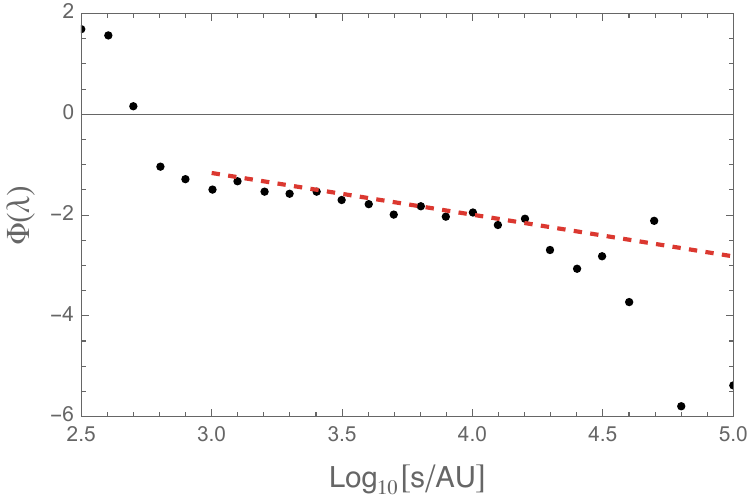}
    \caption{Left panel: The differential function $\Phi(l)$ (Eq. \ref{log.eq}) related to the power index $\gamma(l)$ in the tail of PDF$(a)$ reconstructed by the direct mapping technique. The dashed red line shows the linear fit for the interval $l\in[3,5]$. Right panel: The empirical function $\Phi(\lambda)$ estimated from the actual histogram of $\lambda=\log_{10}(s)$. The dashed red line is the same linear fit as in the left panel, for reference.} 
    \label{dpdl.fig}
\end{figure*}

Using the histogram of reconstructed $l$-values, we can now test the {\"O}pik assumption in the space of true orbital sizes. Indeed, if PDF($a$)$\equiv p(a)$ locally follows the power law 
\eb 
p(a) \propto a^{\gamma(l)},
\label{ag.eq}
\ee 
then from the identity CDF$(\log_{10} a')=$CDF$(a')$ one obtains by differentiating over $a$ the following equation for the PDF of $l$:
\eb 
\Phi(l)\equiv d\,\log_{10}\left(p(l)\right)/dl - 1=\gamma(l)+\frac{d\gamma(l)}{dl}\,l. \label{log.eq}
\ee 
The local value of $\Phi(l)$ in the tail of the reconstructed $a$-distribution can be estimated directly from the histogram $g_i$ of $l$ by computing the decimal logarithm of the histogram values and then the differences between adjacent values divided by the histogram step.
The result for the range $2.5<l<5$ is shown in Fig. \ref{dpdl.fig}, right panel. The bin width (or histogram step) was set to 0.01 to ensure a minimal loss of resolution. The local value of $\gamma$ cannot be directly read from this plot (since there is no exact solution of Eq. \ref{log.eq}), but a fixed model dependence of $\gamma$ on $l$ leads to a definitive parameter estimation problem. $\Phi(l)$ appears to follow a linear function over $l\in[3,5]$, and the dashed red line in the right plot shows the optimal fit
\eb 
\Phi(l)=c_0+c_1\, l
\ee 
obtained by a Nelder-Mead nonlinear optimization with 1-norm merit function. The fitted parameters are $c_0=1.319$, and $c_1=-0.828$. This can now be easily converted into the final approximation for the PDF of $a$:
\eb 
p(a) = K\,a^{b_0+b_1(\log_{10}(a)-3)},
\label{pa.eq}
\ee 
where $K$ is a normalization constant. The values from the direct mapping histogram are $b_0=0.077$ and $b_1=-0.414$.

Eqs. \ref{log.eq} and \ref{pa.eq} were validated by dedicated numerical simulations. A large sample of random $a$-values ($5\cdot 10^5$) was generated with a PDF following Eq. \ref{pa.eq}. A histogram $g_i$ of corresponding $l$-values was computed with a step of 0.1 and the local estimates of $\Phi(l)$ were computed from it. The much larger grid step is needed in this case because of the random number generation noise, which is unchecked by a smoothing kernel. The resulting sequence of $\Phi(l)$ points was almost perfectly consistent with the linear fit shown in Fig. \ref{dpdl.fig}. The values of coefficients $b_0$ and $b_1$ were also verified by performing a direct fit of the function (\ref{pa.eq}) to the histogram of reconstructed $a$-values, producing a practically identical match.

The results obtained by DM indicate that the local power index $\gamma(l)$ is very well approximated by a declining linear fit, which is slightly positive at $l=3$. Furthermore, the power index remains greater than $-1$ everywhere within the data range. The tail of the $a$-distribution is shallow and slowly declining. This result confirms the earlier conclusions based on much smaller and more restricted samples of nearby binaries \citep{1989ApJ...347..998E}. To verify that this conclusion is not a feature of the method, an analogous calculation of $\Phi(\lambda)$ was performed for the observed separations, $\lambda=\log_{10}(s)$. Fig. \ref{dpdl.fig}, right panel, shows the empirical values of $\Phi(\lambda)$ and the same linear fit as in the left panel for $\Phi(l)$. We find that these functions are similar for $a$ and $s$, confirming that the shallow tail is already present in the observed distribution of separations, and the reconstruction procedure does not radically change its form.

Finally, a complete duty cycle simulation was performed to estimate the possible bias in the estimation of $\gamma(l)$ introduced by the smoothing regularization of the reconstruction method. The previously investigated synthetic sample of $a$-values following the PDF with $\gamma(l)=0.077-0.414\,(l-3)$ was converted to a synthetic sample of $s$-values using half a million randomly generated projections $f$. The direct mapping restoration of $a$ was performed on this synthetic $s$-sample and the result was compared with the initial $\gamma(l)$. The resulting $\Phi(l)$ was found to be systematically higher than the initial straight line by 0.204 on average. Thus, the expected dependence should be updated to
\eb 
\gamma^{(DM)}(l)=-0.127-0.414\,(l-3).
\label{gadm.eq}
\ee

\section{Reconstruction of orbit separation by inverse filtering}
\label{ml.sec}
The direct mapping $s\rightarrow a$ employed in Sect. \ref{di.sec} yields the important result that the rate of decrease of probability density with $a$ is slower than the canonical power law with a power index of $-1$. As far as the widest separations are concerned, there may be legitimate concerns about the reliability of the reconstructed PDF because the mapping operation is singular for $f$ tending to $-1$ (i.e., the statistically rare instances of small $s$ corresponding to very large $a$). To verify the DM result, we can solve the inverse problem of reconstructing the distribution of $a$ from the known mapping $a\rightarrow s$ (which is singularity-free) and the given sample distribution of $s$. The problem is reminiscent of the deconvolution problem in image restoration, which is known to be ill-posed in the presence of additive noise and imprecise knowledge of the convolution kernel (response). Solving it requires specific regularization constraints applied to either the 2D Fourier transforms of the data arrays or to iterative nonlinear optimization procedures. In our case, there is no explicit convolution involved in the mapping of unknown $a$ to the given $s$ or back via the random factor $f$. However, the mapping simplifies to a couple of symmetric convolution operators in the $\log$--$\log$ space $\{l,\lambda\}$:
\begin{eqnarray} 
C(l) & = & \int_0^{+\infty} p(\lambda')\,(1-C_F(\lambda'-l))\,d\lambda' \nonumber \\
C(\lambda) & = & \int_0^{+\infty} p(l')\,C_F(\lambda-l')\,dl' ,
\end{eqnarray}
where $F=\log_{10}(1+f)$, $\lambda=\log_{10}(s)$, $C$ is the CDF of the corresponding variate, and $p$ is the PDF.
Upon differentiating these equations, we note that the convolution kernels are mirrored functions $p_F$ around zero value, with asymmetric supports $[-\infty,\log_{10}(2)]$ and $[-\log_{10}(2),+\infty]$, respectively. The infinite support of the kernel makes the numerical implementation of deconvolution practically impossible in the $\log$--$\log$ space.
The principle of the previously employed DM method is based on the concept of maximizing the uncertainty of the possible distribution of $a$ for each specific object with a given separation $s$ that is still consistent with the statistical mapping $a\rightarrow s$. This means in practice that we build up the PDF$(a)$ by the smoothest possible increments that are consistent with the system response, and the result is the smoothest PDF realization of all possible solutions. A legitimate concern is that the reconstructed PDF$(a)$ overestimates the rate of extremely wide pairs, and a dedicated numerical simulation was performed to estimate the bias and correct the $\gamma(l)$ fit in Eq. \ref{gadm.eq}. A viable alternative, which should have the opposite effect, is the inverse filtering (IF) with impulse updates. The increment in the IF method is a delta function, i.e., all the added values $a$ in each iteration have the same value, which corresponds to the modal $s$ value minimizing the chosen merit function. The marginal case is a probability density that is a delta function (all pairs in the sample have the same orbit size), as long as the synthesized $s$-distribution is consistent with the data. The closest analog in the signal deconvolution literature is perhaps Wiener filtering \citep{2022_Weiss}, although the latter is performed in the Fourier domain with an embedded regularization for the known (or assumed) spectral SNR. In our case, the IF approach is realized in the space of binned histograms without any regularization, with the penalty of uncontrolled growth of the input Poisson noise.

The practical realization of the IF method in this study is performed in the space of $g_i(l)$ and $y_i(\lambda)$ histograms on the same dense grid of nodes, with the goal of incrementally building up the working histogram $y(\lambda)$ until it is consistent with the empirical histogram $h(\lambda)$. We start with a flat zero histogram of $g_i$ and compute 4000 random realizations of $1+f$ for eccentricity index $\alpha'=+0.15$. The corresponding sample of possible $\lambda$ values is computed for each nodal value of $l$ by adding $\log_{10}(1+f)$, and the result is converted to histogram increments renormalized to a fixed small number (e.g., $\delta=30$) by the corresponding binning. Then, we select only the single node $l_j$ that provides the minimum GoF metric $\sum_i (h_i-y_i)^2$. The entire increment $\delta$ is added to the single value $g_j$. The iteration cycles until the sum of $y_i$ becomes greater than the sum of $h_i$. Essentially, in comparison to the DM method, the impulse-like and smooth response-function increments are swapped between $g_i(l)$ and $y_i(\lambda)$.

\begin{figure*}
    \includegraphics[width=0.47 \textwidth]{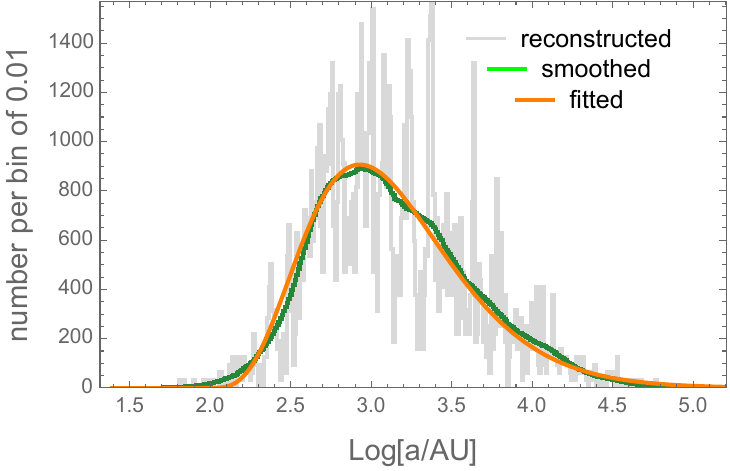}
    \includegraphics[width=0.47 \textwidth]{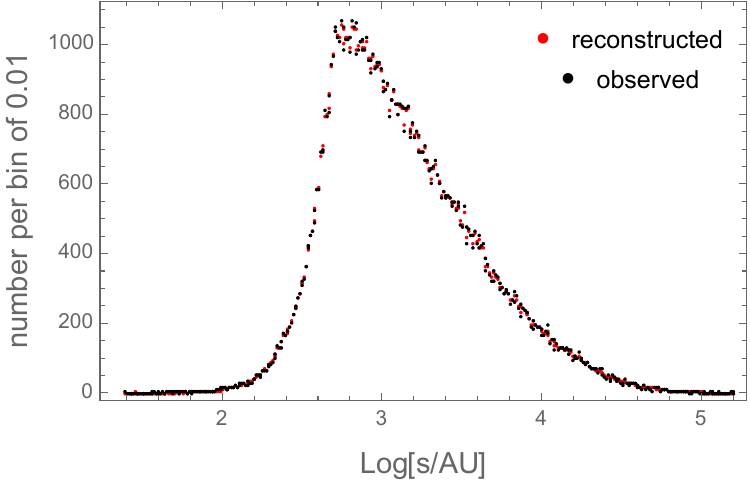}
    \caption{Numerically reconstructed probability density of orbit sizes $a$ and apparent separation $s$ per decibel using the IF method. Left panel: the orange line is a fit per Eq. \ref{gamma2.eq}, the green curve is a kernel smoothing result of the actual reconstructed PDF, which is shown with the gray line. Right panel: the observed (blue) and reconstructed (red) histograms of $\log_{10}(s)$.}
    \label{ml.fig}
\end{figure*}

Fig. \ref{ml.fig}, left panel, shows the result from one such calculation. The IF-reconstructed PDF$(l)$ in the form of $g_i$ histogram is reproduced with a gray line. The main drawback of the IF method is evident in this plot---the shredded character of the histogram reflecting the unabated growth of statistical noise in the data $h_i(\lambda)$. The situation is similar to the deconvolution of blurred images with noise using a high-bandwidth Wiener filter. For a comparison with the previous results, it is practical to use a smoothed version of the reconstructed histogram $g_i(l)$. This can be achieved with two separate smoothing operations: a posteriori Gaussian kernel convolution of the jagged histogram (green curve) or the empirical fit of a Gamma distribution function with free scale and shape parameters (orange curve). Despite the shredded character of the $g_i$ histogram, the corresponding reconstructed histogram $y_i$ faithfully reproduces the given sample distribution $h_i$ of projected separations (Fig. \ref{ml.fig}, right panel). 

The analytical fit for the IF solution is
\eb 
{\rm PDF}(l)=\Gamma[5.73,0.21](l-1.96).
\label{gamma2.eq}
\ee 
In comparison with the DM fit (Eq. \ref{gamma1.eq}), the shape parameter is slightly larger and the scale parameter is slightly smaller. This corresponds to a somewhat wider PDF$(l)$ core for the IF fit, as expected. Consequently, the tail of the distribution is lower for the IF fit, signifying a lower overall rate of very wide binaries. Using the analytical Gamma distribution fit, we estimate the survival function to be 1.2\% at $l=4.5$, 0.23\% at 5.0, and 0.082\% at 5.3. Remarkably, the two methods produce nearly the same rates for extreme orbit sizes $a\gtrsim$ 1 pc. It is noted again that these estimates do not describe the general population of field binaries but only pertain to the given sample.

\begin{figure*}
    \includegraphics[width=0.47 \textwidth]{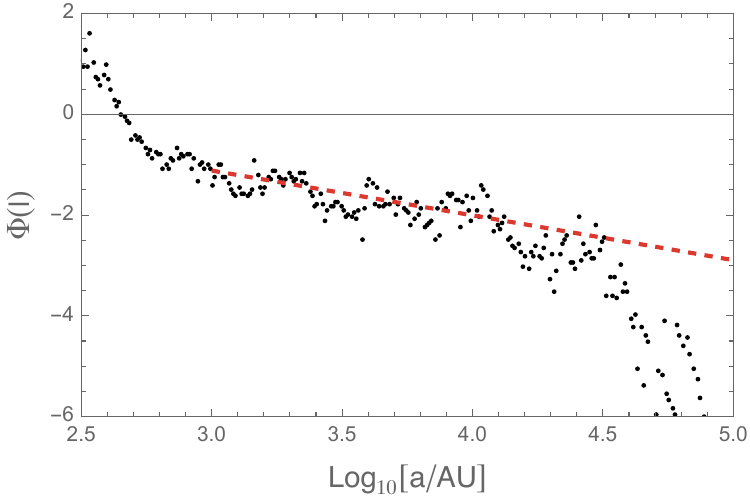}
    \caption{The differential function $\Phi(l)$ (Eq. \ref{log.eq}) related to the power index $\gamma(l)$ in the tail of PDF$(a)$ reconstructed by the IF technique. The dashed red line shows the same linear fit for the interval $l\in[3,5]$ as in Fig. \ref{dpdl.fig}, for reference.} 
    \label{phi_if.fig}
\end{figure*}

The fit in Eq. \ref{gamma2.eq} can be used to roughly estimate the function $\Phi(l)$ in the tail of the IF-reconstructed distribution. This is achieved by computing the derivative of the logarithm of the fit on a grid of points, following Eq. \ref{log.eq}. The resulting curve (not shown for brevity), which is smooth by construction in this case, monotonically declines within the interval $3.0<l<5.0$ confirming the results obtained with the DM in Fig. \ref{dpdl.fig}. However, more useful estimates can be produced by smoothing the reconstructed distribution of $l$ and computing the differentials of its logarithm. Using a Gaussian smoothing kernel with $\sigma_l=0.2$, a grid of $\Phi(l)$ estimates was computed as described in Sect. \ref{di.sec}, and plotted in Fig. \ref{phi_if.fig}.  The linear fit found by the DM method (shown with the red line) is still adequate in the interval $l\in[2.8,4.5]$ but it begins to overestimate the rate of binaries above 4.5 ($a\simeq 32$ KAU).

As a sanity check, a special numerical experiment was performed with a synthetic PDF($a$) representing the marginal case when all sample orbits have the same size $a_0$, i.e., PDF($a$)$=\delta(a_0)$. A random sample of 103169 $f$-values (the size of the data sample) was generated for $\alpha=+0.15$. Input $s$-values for the IF reconstruction were computed by multiplying the $f+1$ vector by $a_0=1000$. The simulated  histogram $h_i$ of $\log_{10}(s)$ (not shown for brevity) is strongly asymmetric with a peak above 3 stretching to $-\infty$ at the low end and abruptly truncated at $3+\log_{10}(2)$ at the high end. In the IF reconstruction $s\rightarrow  a$, a new set of 4000 $f$-values was randomly generated for each of 3284 iterations to ensure its independence of the initial mapping realization. The empirical distribution of $f$, which was used in the iterations, was also derived from a separate generation of 100000 random values.
The reconstructed histogram $g_i$ is consistent with the input delta function with only 7\% of the sample values spilling over to a few adjacent bins around $a_0$. Thus, the IF method works as intended in reconstructing the most compact distributions of $a$ from the infinite set of possible solutions.

\section{Effects of the limited angular resolution}
\label{eff.sec}
The sample distribution of $\lambda=\log_{10}{s}$ shown in 
\ref{ml.fig}, right, is sharp-crested and asymmetric around the peak at approximately 2.8. It is characterized by a steep rise on the lower-$s$ shoulder and a pronounced deficit of small separations. We know from the census of the nearest binaries that the population of binary systems is peaked at significantly shorter separations. This feature emphasizes that the source sample of Gaia-resolved pairs does not adequately represent the general population at shorter separations. The obvious reason is the limited angular resolution of Gaia astrometry, which effectively varies around $1\arcsec$ depending on a number of factors, some of which are not reflected in the published data. The filter on parallax, $\varpi > 4$ mas, is used in this study to mitigate the impact of this deficiency on the general sample representation of the reconstructed $a$. The additional filter $\rho>2\arcsec$ is meant to make the selection more uniform in terms of the orbit size statistics. The combination of these filters implies that any pairs in the source catalog with $s<500$ AU are rejected. This imposed limit corresponds to a hard limit $a<250$ AU. Intuitively, the hard limit should not affect our main conclusions about the relative rate of pairs in the remote tail, but that supposition may benefit from additional verification.

\begin{figure*}
    \includegraphics[width=0.47 \textwidth]{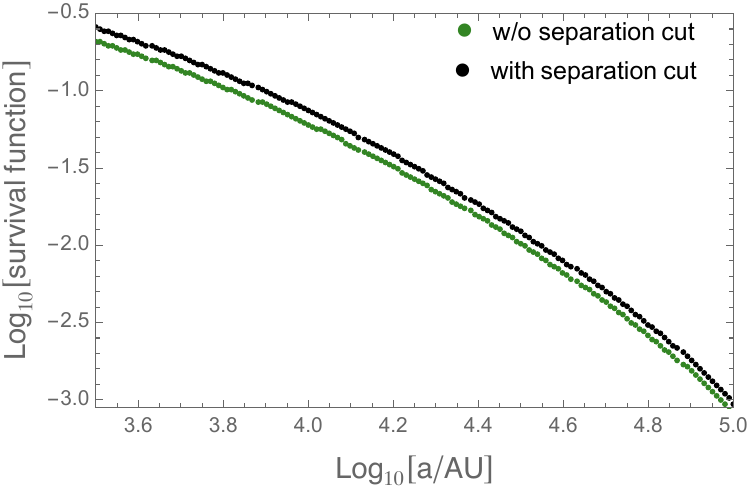}
   \caption{Survival functions ($1-$CDF$(l))$ of the probability density of orbit sizes $l=\log_{10}(a)$ for the widest binaries reconstructed by direct MC mapping for two input sources of resolved pairs. The black curve represents the original smaller sample with a cut at angular separation below $2\arcsec$, while the green curve shows the result for a larger sample without this cut.}
    \label{surv.fig}
\end{figure*}

The procedure of orbit size reconstruction by DM described in Sect. \ref{di.sec} was repeated for a sample with the same parallax and {\tt{ruwe}} conditions but without any imposed limit on angular separation. Consequently, the working sample increased in size from 103,169 to 127,945 pairs. The statistical projection kernel PDF($f$) for $\alpha'=+0.15$ can be used again because it is not sample-specific. The sample distribution of $\lambda$ (not shown for brevity) is less sharp-crested than its counterpart in Fig. \ref{ml.fig}, with the bulk of the additional pairs ending up in the incremented low-end shoulder between 2 and the peak at $\sim 2.7$. The reconstructed $a$-distribution shows a visibly flatter and wider peak with a modal value around 2.6, which cannot be adequately approximated with a Gamma distribution, in contrast with Fig. \ref{di.fig}. The changes in the high-end tail of the distribution of $l$, which are relevant for this study, are depicted in Fig. \ref{surv.fig}. The graph presents the logarithm of the survival function (i.e., $1-\rm{CDF}$) for the two working samples: the larger sample without the separation cut (green dots) and the smaller one with this cut (black dots). As expected, the inclusion of tighter binaries makes the survival function generally lower in the extended tail, but the difference in the wide dynamic range $\l\in [3.5,5.0]$ emerges as a rather uniform scale factor. This experiment shows that the shape of the extended tail is little sensitive to the imposed cuts on angular resolution.

The numerical experiment described in Sect. \ref{ml.sec} with an initial PDF($a$)$=\delta(1000)$ was further extended to estimate the impact of the limited angular resolution on the available sample of systems. The randomly generated sample of projected separations $s$ was decimated according to the imposed limit $\rho>2\arcsec$ by assigning an observed parallax value to each synthetic $s$ (without replacement) and computing the resulting angular separations. The censored sample, which was reduced in size from 103169 to 93508, was processed in the same way as described above using the IF restoration method. Without any modification of the $s\rightarrow  a$ kernel for censoring, the restored $a$-distribution is a sharp peak with nonzero values concentrated in 5 adjacent bins of 0.01 width. The effect of sample censoring is seen in the slightly asymmetric shape of the peak leaning toward $l$ above 3.

\section{Conclusions}
\label{end.sec}

Two important and currently poorly understood physical processes are imprinted in the distribution of field gravitationally bound stellar systems: the origin of wide binaries and the survival of low-energy, weakly bound pairs at the early cluster stages and their more protracted dynamical evolution and chance interactions with other stars and clouds in the field. Unfortunately, these two effects are entangled and coupled in the present-day statistics, so they cannot be easily separated. One possible approach that has been explored in the literature is to analyze possible differences in the rates of wide apparent separations for dynamically distinct Galactic populations, e.g., the disk and the halo groups. One of the most recent works by \citet{Tian_2020} finds approximately the same spectral power index for different kinematic populations at $s<10^4$ AU, but a steeper decline for the tentative halo population. This contradicts the initial expectation that wide pairs in the disk are disrupted sooner via interaction with the denser environment (molecular clouds and other stars). The uniformity of the power-law index for the empirical distribution of angular separations in the halo has been used to put stringent constraints on the contribution of invisible MACHOs to the dark matter mass \citep{2004ApJ...601..311Y}. Alternative forms of dark matter, such as spatially extended solitons, can be invoked to explain the apparently enhanced disruption of the widest binaries \citep{2025JCAP...02..001Q}.

We note that these previous case studies have been mostly based on the distribution of apparent separations without taking into account the effects of statistical projection and eccentricity. The reconstructed probability distributions in the space of true semimajor axes $a$ reveal a slower power-law decline 
than the often assumed $\gamma=-1$ for $a$ up to 50--100 KAU (Figs. \ref{dpdl.fig} and \ref{phi_if.fig}). Generally, the tail of the PDF$[a]$ is slightly heavier than the tail of PDF$[s]$ (Fig. \ref{dpdl.fig}). The reason for this difference is in the asymmetric character of the factor $1+f$ in the mapping $a\rightarrow s$ (Fig. \ref{di.fig}), i.e., the fact that the projection can make the separation $s$ much smaller than $a$ (down to zero), while the eccentricity effect can make $s$ only somewhat larger than $a$ (up to a factor of 2). The theoretical studies of the disruption process in the impulsive and tidal approximations \citep{1987ApJ...312..367W} concern the population distribution of orbital axes (and orbital periods), and we have provided in this paper information about the rates of wide systems in the tail of the distribution.

We find a finite number of extremely wide binaries in the field with semimajor axes stretching toward 1 pc. From the analysis of the Gaia-detected sample of wide resolved binaries with two conceptually different methods, we find evidence of a continuous tail in the distribution of $a$ without a specific cutoff value. This result is consistent with the historically early conclusions drawn from much smaller samples of nearby stars \citep{1987ApJ...312..390W}. Although the rate of extreme orbit sizes approaching 1 pc is indeed quite small, the presence of such systems poses significant challenges for the models of dynamical disassociation in the Galactic field. Even a single flyby of a field star at a sufficiently small impact parameter may disrupt a system with a low escape velocity via an impulse-like increase of a component's velocity \citep{1975MNRAS.173..729H}. Sufficiently old binaries in the disk should have survived multiple interactions of that kind. A hidden mass in a hierarchical multiple system is a possible way out of this conundrum. The observable manifestation of a dim massive companion (such as a stellar-mass black hole) is a large and discrepant proper motion of the observed companions. However, the consistency criterion for proper motion used in the creation of the clean sample should have removed most such cases in our collection of nearby objects. 

The tidal gradient of the local Galactic potential can also be sufficient to disassociate a wide stellar binary \citep{2010MNRAS.401..977J}. Some potentially testable factors contributing to this mechanism are the total mass of the system, the mean radius of the  Galactic orbit, and the effective elongation of the Galactic orbit. It was shown that the orientation of the binary momentum with respect to the Galactic orbital momentum is also a critical factor \citep{2012MNRAS.421L..11M}. Marginally soft systems are much more stable in the Galactic field if their orbital momentum is counter-directed to the Galactic rotational spin. Furthermore, a counter process of stochastic capture of unrelated stars into long-term stable pairs with retrograde motion is also possible. The widest pairs detected by Gaia may then be mostly retrograde and transient associations. This hypothesis is testable with high-accuracy radial velocity measurements of the widest candidate binaries.

Alternative models of gravitation in the weak-field regime, such as MOND, provide attractive possibilities too.
The observational difficulties related to the sensitivity of relative proper motions to multiplicity and measurement error are significant in this context \citep{2023OJAp....6E...2M, 2023OJAp....6E...4P}. Both the long-term stability and the probability of capture of unrelated stars are drastically enhanced in the MOND regime \citep{2012MNRAS.421L..11M}. Taking into account the external field effect from the ambient Galactic potential, MOND-like orbits are tighter at  high values of eccentricity, while the orbital motion is faster than that in the Newtonian regime, and the osculating elements (including eccentricity and the line of apsides) rapidly change on the scale of one orbital period \citep{2010OAJ.....3..156I}. Thus, the boosted acceleration in the framework of MOND should increase the rate of very wide binaries, increase the relative velocity of the companions at a fixed separation, and reshape the distribution of eccentricity. For a total mass of $1\,M_\sun$, the characteristic distance where the Newtonian gravitational acceleration becomes equal to the critical acceleration in MOND is 7 KAU. However, employing the smooth interpolation function for acceleration from \citet{2005MNRAS.363..603F}, we estimate that a significant 1\% boost of orbital acceleration is achieved at a much shorter separation of $\sim 700$ AU. The effects of modified gravity may be imprinted in the parameters of a large fraction of the Gaia-resolved binary systems.

Our analysis is still limited by the overall sample statistics. For an object-specific analysis, we can only estimate the probability of $a$ to be in a certain interval from a given $s$ using the probability densities in Fig. \ref{f.fig} for a fixed eccentricity index $\alpha'$. This information can be used in a forward modeling of the distribution of total orbital energy in different kinematic populations (e.g., disk binaries versus halo binaries) by estimating the mass of the components from stellar evolution models or directly from the derived Gaia data. Additional inroads into the theory of dynamical evolution of weakly bound systems can be achieved by correlating the statistical expectation of orbital size or orbital energy with the vertical component of the observed systemic velocity in the Galactic coordinate system. Binary systems that travel at high velocity through the Galactic disk (and subsequently spend most of their time in the low-density environment outside the disk) are expected to be wider and softer on average.

\section*{Acknowledgements}
The author thanks the referee A. Tokovinin for thoughtful and deep reviews of this paper, which resulted in substantial improvements and corrections.

\bibliography{main}
\bibliographystyle{aasjournal}

\end{document}